\documentclass{article} 

\usepackage{gen2_iclr2026,times}

\usepackage{amsmath,amsfonts,bm}

\def\eqref#1{equation~\ref{#1}}

\def\1{\bm{1}}

\DeclareMathAlphabet{\mathsfit}{\encodingdefault}{\sfdefault}{m}{sl}
\SetMathAlphabet{\mathsfit}{bold}{\encodingdefault}{\sfdefault}{bx}{n}

\usepackage{hyperref}
\usepackage{url}
\usepackage{booktabs}
\usepackage{amsmath}
\usepackage{amsfonts}
\usepackage{amssymb}
\usepackage{graphicx}
\usepackage{xcolor}
\usepackage{algorithm}
\usepackage{algorithmic}
\usepackage{multirow}
\usepackage{subcaption}
\usepackage{wrapfig}
\usepackage{microtype}
\usepackage{longtable}
\usepackage{array}
\usepackage{makecell}

\newcommand{\geneval}{\textit{GGE}}

\newcommand{\mmd}{\text{MMD}}
\newcommand{\reals}{\mathbb{R}}

\title{A Standardized Framework for Evaluating Gene Expression Generative Models}

\title{A Standardized Framework for Evaluating Gene Expression Generative Models}

\author{
Andrea Rubbi$^{1,2,\dagger}$ \qquad
Andrea G. Di Francesco$^{3,4,\dagger}$ \qquad
Mohammad Lotfollahi$^{2,5,6}$ \qquad
Pietro Li\`o$^{1}$\\
\small
\\
$^{\dagger}$Equal contribution.\\
$^{1}$Department of Computer Science and Technology, University of Cambridge, Cambridge, United Kingdom\\
$^{2}$Wellcome Sanger Institute, Cambridge, United Kingdom\\
$^{3}$Sapienza University of Rome, Rome, Italy\\
$^{4}$ISTI-CNR, Institute of Information Science and Technologies, Pisa, Italy\\
$^{5}$Cambridge Centre for AI in Medicine, University of Cambridge, Cambridge, United Kingdom\\
$^{6}$Cambridge Stem Cell Institute, University of Cambridge, Cambridge, United Kingdom
}

\iclrfinalcopy 

\begin{document}

\maketitle

\begin{abstract}
The rapid development of generative models for single-cell gene expression data has created an urgent need for standardised evaluation frameworks. Current evaluation practices suffer from inconsistent metric implementations, incomparable hyperparameter choices, and a lack of biologically-grounded metrics. We present \textit{Generated Genetic Expression Evaluator} (\geneval{}), an open-source Python framework that addresses these challenges by providing a comprehensive suite of distributional metrics with explicit computation space options and biologically-motivated evaluation through differentially expressed gene (DEG)-focused analysis and perturbation-effect correlation, enabling standardized reporting and reproducible benchmarking. Through extensive analysis of the single-cell generative modeling literature, we identify that no standardized evaluation protocol exists. Methods report incomparable metrics computed in different spaces with different hyperparameters. We demonstrate that metric values vary substantially depending on implementation choices, highlighting the critical need for standardization. \geneval{} enables fair comparison across generative approaches and accelerates progress in perturbation response prediction, cellular identity modeling, and counterfactual inference.

\end{abstract}

\begin{figure}[h]
    \centering
    \includegraphics[width=0.8\linewidth]{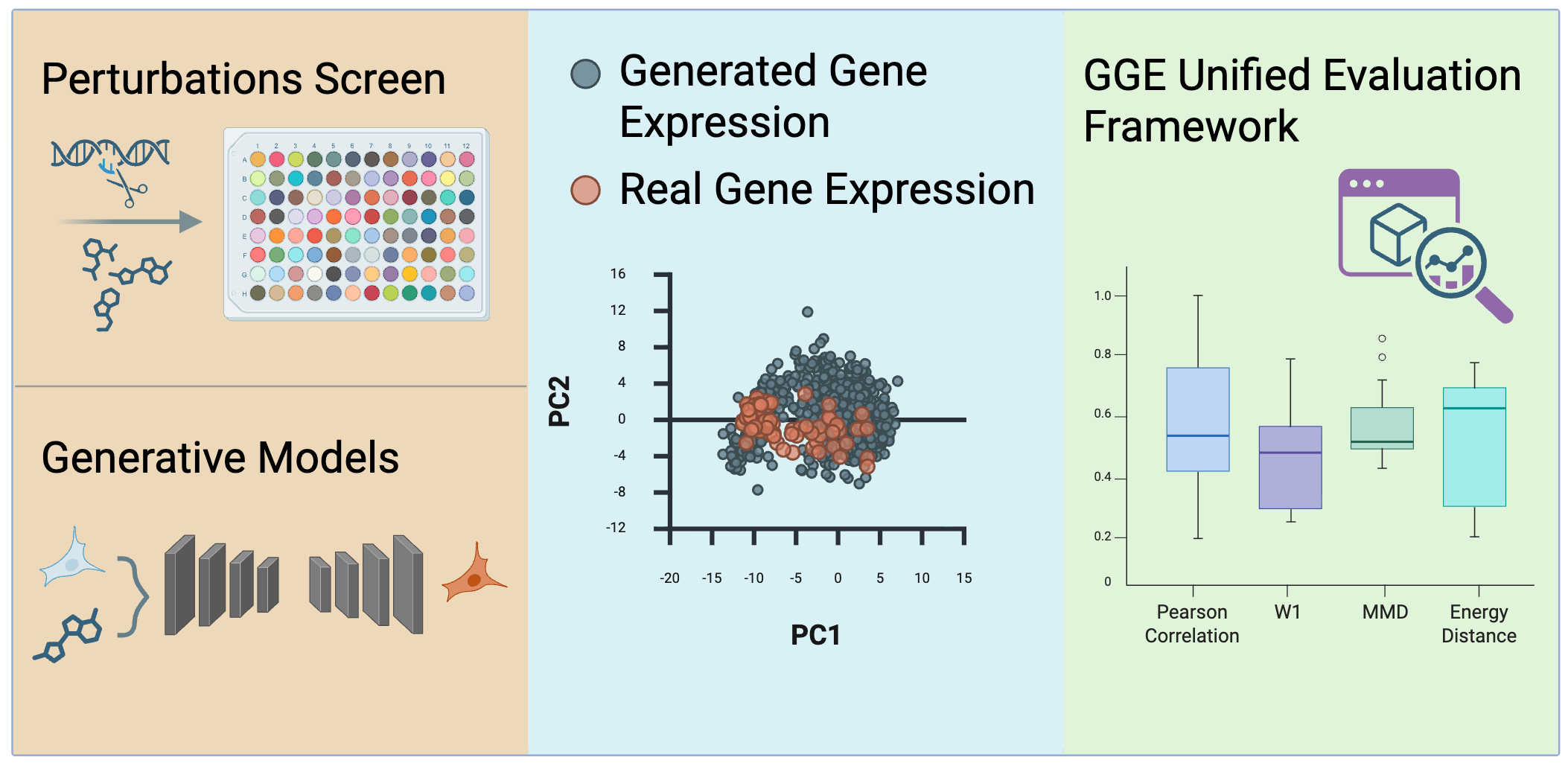}
    \caption{GGE is a unified framework for the evaluation of generated genetic expression profiles. GGE aims to standardise the otherwise heterogeneous set of evaluation protocols proposed by the multitude of generative models implemented for this task.}
    \label{fig:gge_framework}
\end{figure}

\section{Introduction}

Generative models for single-cell gene expression data have emerged as powerful tools for understanding cellular biology, with applications spanning perturbation response prediction, developmental trajectory modeling, and counterfactual inference for therapeutic discovery. The field has witnessed remarkable progress: variational autoencoders (VAEs) established early benchmarks for perturbation prediction \citep{lotfollahi2019scgen}, graph neural networks enabled reasoning over gene regulatory networks \citep{roohani2024predicting}, and optimal transport methods provided principled frameworks for modeling cellular transformations \citep{bunne2023learning}. Most recently, flow matching approaches have demonstrated state-of-the-art performance on distribution matching tasks \citep{lipman2022flow, tong2023conditional}, opening new possibilities for learning continuous transformations between cellular states.

Despite this rapid progress, a fundamental problem persists: the field lacks standardized evaluation practices. This absence of standardization has profound consequences for scientific progress. When different methods report ``Wasserstein distance'' using incompatible formulations--per-gene 1D distances averaged across genes, multivariate distances in raw $\reals^{2000}$ space, or distances in PCA-reduced $\reals^{50}$ space--meaningful comparison becomes impossible. When critical hyperparameters such as Sinkhorn regularization strength, kernel bandwidth for MMD, or DEG significance thresholds go unreported, results cannot be reproduced or compared. When aggregate metrics over all genes obscure biologically relevant differences concentrated in a small subset of differentially expressed genes, we may miss the most important signals.

The heterogeneity of evaluation practices in the current literature is striking. We surveyed 12 influential methods for single-cell generative modeling (Table~\ref{tab:metrics_survey}) and found that no two papers use identical evaluation protocols. Some methods report only reconstruction metrics (MSE, $R^2$), while others focus on distributional metrics; some compute metrics in raw gene space, others in PCA space; some evaluate on all genes, others on DEGs alone. This inconsistency makes it virtually impossible to determine which methods represent genuine advances.

We present \geneval{}, an open-source Python framework that addresses these challenges through two key design principles. First, we make all implementation choices explicit through a unified API with a \texttt{space} parameter (``raw'', ``pca'', or ``deg'') and standardised parameters. Second, we enable biologically meaningful evaluation through perturbation-effect correlation computed in DEG space, measuring whether models capture the direction and magnitude of perturbation effects rather than merely reconstructing expression levels. Our goal in this work is to standardize and expose evaluation choices that are often left implicit, rather than to claim a definitive benchmark across all model families.

\section{Related Work: The Evaluation Landscape}

\subsection{Generative models for single-cell perturbation data}
Generative modeling for single-cell gene expression has progressed from VAE-based likelihood models to distribution-matching and foundation-model approaches.
Early work such as scVI~\citep{lopez2018deep} and scGen~\citep{lotfollahi2019scgen}
established VAE baselines for denoising and perturbation prediction, later extended to combinatorial perturbations by CPA~\citep{lotfollahi2021compositional}. Biology-informed architectures incorporate prior knowledge, e.g. GEARS~\citep{roohani2024predicting} injects gene--gene interaction structure to improve generalization to unseen perturbations.
In parallel, optimal transport formulations (e.g.\ CellOT~\citep{bunne2023learning})
and flow matching methods~\citep{lipman2022flow,tong2023conditional,klein2025cellflow, rubbi2026mixflow}
explicitly target \emph{conditional distributions} rather than conditional means.
Finally, transformer-based foundation models (e.g.\ STATE~\citep{adduri2025state}) emphasize
large-scale representation learning and multi-task generalization across contexts.
A more detailed landscape review is provided in Appendix~\ref{app:gen_models_landscape}.

\subsection{Survey of Evaluation Practices}

To quantify the heterogeneity of evaluation practices in the literature, we surveyed 12 influential methods for single-cell generative modeling. Table~\ref{tab:metrics_survey} summarizes the metrics, computation spaces, and evaluation protocols used by each method. Several patterns emerge from this analysis.

\begin{table}[t]
    \centering
    \caption{Survey of evaluation metrics used by single-cell generative models. ``Space'' indicates whether metrics are computed in raw gene space (Raw), PCA space (PCA-$k$), or on highly variable genes (HVG). Dashes indicate the metric was not reported. Recent methods (STATE, CellFlow) show a trend toward more comprehensive evaluation, yet protocols remain inconsistent.}
    \label{tab:metrics_survey}
    \vspace{0.5em}
    \resizebox{\textwidth}{!}{
        \begin{tabular}{@{}lccccccl@{}}
            \toprule
            \textbf{Method} & \textbf{MSE} & $\boldsymbol{R^2}$ & $\boldsymbol{W_1/W_2}$ & \textbf{MMD} & \textbf{Corr} & \textbf{DEG-specific} &  \textbf{Space} \\
            \midrule
            scVI \citep{lopez2018deep} & \checkmark & \checkmark & \textemdash{} & \textemdash{} & \textemdash{} & \textemdash{} & Raw \\
            scGen \citep{lotfollahi2019scgen} & \checkmark & \checkmark & \checkmark & \checkmark & \checkmark & Top-100 DEG & Raw \\
            Waddington-OT \citep{schiebinger2019optimal}  & \textemdash{} & \textemdash{} & \checkmark & \textemdash{} & \textemdash{} & \textemdash{} & PCA-30 \\
            CellOT \citep{bunne2023learning} & \checkmark & \textemdash{} & \checkmark & \textemdash{} & \textemdash{} & \textemdash{} & PCA-50 \\
            CPA \citep{lotfollahi2021compositional} & \checkmark & \checkmark & \textemdash{} & \textemdash{} & \checkmark & Top-50 DEG & Raw \\
            GEARS \citep{roohani2024predicting} & \checkmark & \textemdash{} & \textemdash{} & \textemdash{} & \checkmark & Top-20 DEG & Raw/HVG \\
            scVelo \citep{bergen2020generalizing} & \textemdash{} & \checkmark & \textemdash{} & \textemdash{} & \checkmark & \textemdash{} & Raw \\
            PRESCIENT \citep{yeo2021generative} & \textemdash{} & \textemdash{} & \checkmark & \textemdash{} & \textemdash{} & \textemdash{} & PCA-100 \\
            scDiff \citep{tang2023conditionaldiffusion} & \checkmark & \textemdash{} & \checkmark & \checkmark & \textemdash{} & \textemdash{} & PCA-50 \\
            CFGen \citep{palma2025cfgen} & \textemdash{} & \textemdash{} & \checkmark & \checkmark & \textemdash{} & \textemdash{} & PCA-50 \\
            STATE \citep{adduri2025state} & \checkmark & \checkmark & \checkmark & \checkmark & \checkmark & Top-20 DEG & PCA-50 \\
            CellFlow \citep{klein2025cellflow} & \checkmark & \textemdash{} & \checkmark & \checkmark & \checkmark & \textemdash{} & PCA-50 \\
            \bottomrule
        \end{tabular}
    }
\end{table}

First, there is no consensus on which metrics to report. While MSE or $R^2$ appears in 10/12 surveyed methods, distributional metrics ($W_1$, $W_2$, or MMD) appear in 8/12, and correlation-based metrics in 7/12. This heterogeneity makes direct cross-paper comparison difficult, as methods may appear superior by optimizing for mean accuracy while failing to capture the full distributional variance.

Second, implementation details for the same metric name differ substantially. ``Wasserstein distance'' is used inconsistently across the literature, referring variously to the 1D per-gene average $W_1$, a multivariate Sinkhorn approximation, or exact $W_2$ via linear programming on cell subsets. Furthermore, the number of PCA components used as the coordinate space varies from 30 to 100, which directly alters the sensitivity of distance-based evaluations.

Third, biological context is handled inconsistently. While early methods evaluated on all genes, newer perturbation models like GEARS focus on Pearson correlation and MSE calculated specifically on the top 20 differentially expressed genes (DEGs). This shift highlights the importance of capturing the perturbation signal itself rather than just the steady-state transcriptomic background.

These observations motivate the design of \geneval{}: a framework that makes these choices explicit, provides standardized implementations of multivariate and per-gene metrics, and enables fair comparison across disparate modeling architectures.

\section{Theoretical Foundations}
\label{sec:theory}

Before describing \geneval{}'s implementation, we establish the theoretical foundations of distributional metrics for gene expression data and analyze when different computation spaces are appropriate. Our goal is to formalize evaluation choices that are often treated implicitly in prior work, and to clarify their biological interpretation.

\subsection{Problem Formulation}

Let $P = \{x_1, \ldots, x_n\}$ and $Q = \{y_1, \ldots, y_m\}$ be samples from two distributions representing real and generated gene expression data, where $x_i, y_j \in \reals^G$ (expression of $G$ genes). We seek metrics $d(P, Q)$ that quantify distributional similarity. In the context of single-cell perturbation modeling, these distributions represent empirical samples from conditional data-generating processes (e.g.\ control vs.\ perturbed cells), and the choice of metric encodes which biological properties of the distribution we prioritize: reconstruction fidelity, preservation of population heterogeneity, recovery of multimodality, or consistency of gene--gene dependency structure \citep{lopez2018deep, bunne2023learning, lipman2022flow}.

Importantly, unlike bulk RNA-seq, single-cell data exhibits substantial intrinsic stochasticity, cell-state heterogeneity, and technical noise \citep{svensson2017power_scRNAseq}. As a result, evaluating only pointwise errors (e.g.\ MSE between predicted and observed means) conflates distributional mismatch with biological variability. Distributional metrics are therefore essential for assessing whether generative models capture the full conditional distribution of cellular responses rather than only their average behavior \citep{bunne2023learning, tong2023conditional}.

\subsection{Distributional metrics: Optimal Transport, MMD, and Energy}

Distributional metrics are essential for evaluating single-cell generative models because
they assess whether a model recovers the \emph{population-level response} to a perturbation,
including heterogeneity and multimodality, rather than only matching mean expression levels.
This is particularly important in perturbation settings, where interventions often induce
continuous phenotypic shifts and heterogeneous outcomes across subpopulations.

\textbf{Optimal transport (OT) metrics.}
Wasserstein distances (e.g., 1- and 2-Wasserstein) define a geometric notion of discrepancy between probability distributions by measuring the minimum cost required to transport mass from generated samples to real data under a chosen ground metric~\citep{villani2008optimal, peyre2019computational}. 
In biological applications, OT admits a natural interpretation as the minimal displacement of cell populations in expression space required to align perturbed and reference distributions, making it well-suited for modeling cellular state transitions. 
Recent generative modeling frameworks, including flow matching and OT-based approaches, explicitly optimize objectives related to Wasserstein distances, further motivating their use as evaluation criteria~\citep{arjovsky2017wasserstein, lipman2022flow}.

\textbf{Kernel- and distance-based two-sample metrics.}
Maximum Mean Discrepancy (MMD) and Energy Distance provide complementary, nonparametric two-sample statistics for comparing generated and real distributions directly from samples, without assuming parametric density models~\citep{gretton2012kernel, szekely2013energy}. 
Unlike OT, these metrics do not induce an explicit notion of mass transport or sample coupling and therefore lack a direct geometric interpretation in data space. 
However, they are typically more computationally efficient and stable to estimate, which has led to their widespread adoption as benchmarking metrics for generative models.

In \geneval{}, we support OT-, MMD-, and Energy-based metrics under a unified API. Crucially,
we expose all implementation choices (e.g.\ entropic regularization for OT, kernel bandwidth
for MMD) and the representation space in which distances are computed, as these choices
substantially affect reported values (formal definitions and computational details are
provided in Appendix~\ref{app:metrics_theory}).

\subsection{The Space Question: A Theoretical Analysis}
\label{sec:space_theory}

A critical choice in computing any distributional metric is the space in which distances are computed. This choice substantially affects both statistical behavior and biological interpretability \citep{bunne2023learning, lotfollahi2021compositional}.

\textbf{Raw gene space.} Computing distances in $\reals^G$ with $G \sim 5{,}000$--20{,}000 ($G \sim 2{,}000$--5{,}000 when only highly variable genes are selected) preserves gene-level interpretability, allowing direct attribution of discrepancies to specific genes. However, high-dimensional concentration of measure implies that pairwise distances become less discriminative as $G$ grows \citep{beyer1999nearestneighbor}, and noisy lowly expressed genes contribute equally to the metric despite limited biological relevance. Moreover, technical variation and dropout inflate distances in raw space, potentially dominating biological signal \citep{svensson2017power_scRNAseq}.

\textbf{PCA space.} Projecting to $\reals^k$ using PCA mitigates both statistical and computational issues by restricting attention to dominant axes of variation \citep{pearson1901linesplanes}. In single-cell analysis, PCA is routinely used to denoise expression data and capture major biological programs (e.g.\ cell cycle, differentiation trajectories) \citep{luecken2019scrna_tutorial}. However, PCA bases are fit globally and may underrepresent perturbation-specific gene programs that exhibit low variance in control conditions. As a result, PCA-based distances may systematically underweight rare but biologically important responses.

\textbf{DEG-restricted space.} Restricting evaluation to differentially expressed genes focuses metrics on biologically salient perturbation effects, aligning evaluation with common biological validation practices in perturbation studies \citep{replogle2022perturbseq}. However, DEG selection introduces additional hyperparameters (log-fold-change and statistical significance thresholds), and DEG sets may be unstable under small sample sizes or noisy conditions, leading to high variance in downstream metrics.

\paragraph{Data normalization and transformation.}
In addition to the choice of computation space, the preprocessing applied to gene expression values substantially affects metric values. Single-cell RNA-seq data are typically processed through several transformations prior to downstream analysis, including library-size normalization (e.g.\ counts per million or size-factor normalization) and variance-stabilizing transformations such as $\log(1 + x)$ \citep{luecken2019scrna_tutorial}. These transformations alter the geometry of the data: raw count space is dominated by library-size variation and highly expressed genes, whereas normalized and log-transformed representations compress dynamic range and emphasize relative expression differences across genes. Consequently, distributional metrics such as Wasserstein distance, MMD, and Energy distance may yield substantially different values depending on whether they are computed on raw counts, normalized counts, or log-transformed expression matrices.

In practice, most single-cell generative models operate on normalized and $\log(1+x)$-transformed data during training and evaluation, and benchmarking studies often report metrics in this transformed space. However, this choice is rarely documented explicitly in published evaluation protocols, further complicating cross-paper comparison. For reproducibility, GGE assumes that input datasets are provided in the same representation used for model training (typically library-size-normalized and log-transformed expression values) and treats this preprocessing stage as part of the evaluation configuration that should be reported explicitly alongside the computation space.

\paragraph{Biological interpretation of distributional metrics.}  
From a biological standpoint, distributional metrics answer a different question than pointwise errors: they quantify whether a generative model recovers the \emph{population-level response} to a perturbation, including heterogeneity, multimodality, and the relative abundance of subpopulations. This is crucial in applications such as drug response modeling, CRISPR perturbation screens, and cell fate prediction, where interventions often induce heterogeneous outcomes rather than uniform shifts \citep{bunne2023learning, lotfollahi2021compositional}. Evaluating only mean expression risks rewarding models that collapse diverse responses into averaged, biologically implausible predictions.

\paragraph{An effective framework.}  
We recommend a multi-space evaluation strategy: PCA-50 for primary distributional metrics (statistical robustness and computational tractability), DEG-restricted space for biologically targeted evaluation, and raw gene space only when gene-level interpretability is explicitly required. This triangulation reflects the complementary roles of statistical fidelity and biological relevance in benchmarking generative models for single-cell perturbation prediction.

\section{The GGE Framework}

Building on the theoretical foundations established above, we now describe \geneval{}'s architecture, API design, and key innovations.

\subsection{Architecture and Design Principles}

\geneval{} is built around two core design principles that address the standardization challenges identified in our literature survey.

\paragraph{Explicit Configuration.} Every implementation choice that affects metric values is exposed as an explicit parameter. The \texttt{space} parameter (``raw'', ``pca'', or ``deg'') determines the computation space. For PCA space, \texttt{n\_components} specifies the dimensionality. For DEG space, \texttt{deg\_lfc} and \texttt{deg\_pval} control differential expression thresholds, while the \texttt{n\_top\_degs} parameter enables top-N DEG selection to match protocols from papers like scGen (top-100) or GEARS (top-20). For Sinkhorn approximations, \texttt{blur} sets the regularization strength. This explicitness enables reproducibility and fair comparison.

\paragraph{Universal Space Support.} All metrics---distributional, correlation, and reconstruction---support all three computation spaces through a unified interface. A single call can evaluate in multiple spaces simultaneously, generating a comprehensive evaluation report. Internally, \geneval{} manages PCA fitting, DEG computation, and space transformations transparently.

\subsection{API Design}
The core \geneval{} API makes the choice of computation space an explicit part of metric specification, while preserving a simple, unified interface for end users. Each metric is instantiated together with the representation space in which it is evaluated (e.g.\ PCA space for distributional metrics or DEG-restricted space for correlation-based metrics), along with any method-specific hyperparameters such as the number of principal components or DEG selection thresholds.

Evaluation is performed through a single high-level entry point that loads real and generated data lazily, computes metrics independently for each experimental condition (e.g.\ cell type and perturbation), and aggregates results across the dataset. This design enables stratified analysis of model performance across biological contexts while avoiding unnecessary memory overhead for large single-cell datasets. The API returns both per-condition scores and dataset-level summary statistics (mean and standard deviation across conditions), facilitating systematic comparison of generative models under heterogeneous perturbation regimes.

Concrete usage examples illustrating typical evaluation configurations are provided in the official documentation (\href{https://docs.andrearubbi.com/gge}{https://docs.andrearubbi.com/gge}).

\subsection{DEG-Space Evaluation with Perturbation Effects}

For perturbation prediction tasks, a subtle but critical issue arises when computing correlation metrics on expression means. If the control and perturbed conditions have similar mean expression levels, correlation on raw expression will be artificially high regardless of whether the model captures perturbation effects. This occurs because the correlation is dominated by genes that are similarly expressed across conditions, obscuring differences in the genes that actually respond to perturbation.

\geneval{} addresses this through \textit{perturbation-effect correlation}. Instead of correlating raw expression means, we compute:
\begin{equation}
    \rho_{\text{effect}} = \text{corr}(\mu_{\text{real}} - \mu_{\text{ctrl}}, \mu_{\text{gen}} - \mu_{\text{ctrl}})
\end{equation}
where $\mu_{\text{real}}$ is the mean expression of real perturbed cells, $\mu_{\text{gen}}$ is the mean expression of generated cells, and $\mu_{\text{ctrl}}$ is the mean expression of matched control cells. This measures whether the model captures the \textit{direction and magnitude} of perturbation effects---the signal of biological interest.

When using DEG-space correlation, \geneval{} first identifies DEGs between control and perturbed conditions using a Wilcoxon rank-sum test (or alternative methods), then computes perturbation-effect correlation on the subset of significant genes. This combination focuses evaluation on the biologically relevant signal while properly measuring perturbation effects.

\subsection{Condition-Aware Evaluation}

Single-cell perturbation datasets typically contain multiple cell types and perturbation conditions, each potentially exhibiting distinct DEG sets and response magnitudes. \geneval{} evaluates metrics per condition (cell type $\times$ perturbation pair), enabling stratified analysis. This design choice surfaces heterogeneity across conditions that aggregate metrics would obscure.

For DEG-space metrics, this condition-aware design is essential: DEGs are computed separately for each condition based on comparison to matched control cells. A gene may be a DEG for perturbation A in cell type X but not for perturbation B in cell type Y. \geneval{} automatically handles this complexity, computing condition-specific DEG sets and applying them to the appropriate subsets of data.

\section{Experiments}
\label{sec:experiments}

We present experiments demonstrating the importance of standardization and showcasing \geneval{}'s capabilities across multiple benchmarks.

\subsection{The Importance of Standardization}

Our central claim is that metric values depend critically on implementation choices. To demonstrate this concretely, we compute distributional metrics on identical data under varying configurations. We benchmark on the Norman dataset (39k cells $\times$ 2000 genes in the test set, 138 perturbation conditions) against generated data comprising 1,000 cells per condition (138,000 cells in total) using a recent flow-matching model (MixFlow \citep{rubbi2026mixflow}), then systematically vary the computation space and measure resulting metric values.

\begin{table}[h]
\centering
\caption{Distributional metric values under different computation spaces. The same data produces $\sim$5--10$\times$ different metric values depending on space and dimensionality. All metrics computed using \geneval{} v0.6.2 defaults on MPS(M3).}
\begin{tabular}{lccc}
\toprule
Configuration & $W_2$ & Energy & Time (s) \\
\midrule
Raw ($G = 2000$)   & $104.3 \pm 0.3$ & $0.65 \pm 0.01$ & 1.9 \\
PCA-100            & $53.8 \pm 0.1$  & $0.20 \pm 0.00$ & 3.0 \\
PCA-50             & $33.6 \pm 0.1$  & $0.09 \pm 0.00$ & 2.7 \\
PCA-25             & $17.2 \pm 0.1$  & $0.09 \pm 0.00$ & 2.4 \\
\bottomrule
\end{tabular}
\label{tab:space_comparison}
\end{table}

Table~\ref{tab:space_comparison} reports remarkable, yet expected, differences: $W_2$ distance varies by nearly an order of magnitude (17.2 to 104.3) depending solely on computation space. This variation is not random noise---it reflects the mathematical reality that distance metrics scale with dimensionality and feature selection. A paper reporting ``$W_2 = 17.2$'' computed in PCA-50 space cannot be compared to another reporting ``$W_2 = 104.3$'' in raw space, yet both might be described simply as ``Wasserstein distance'' without further specification.

\subsection{Ablation: Effect of DEG Thresholds}

For DEG-space metrics, the choice of significance thresholds affects both the number of genes included and the resulting metric values. \geneval{} supports both threshold-based DEG selection and top-N selection (e.g., top-100 DEGs as used by scGen, or top-20 as used by GEARS).

\begin{table}[h]
\centering
\caption{Effect of DEG selection strategy on correlation metrics. Top-N selection provides consistent gene counts across conditions; threshold-based selection adapts to perturbation strength. We benchmark on the Norman dataset (39k cells $\times$ 5000 genes) with 138 perturbation conditions.}
\label{tab:deg_thresholds}
\begin{tabular}{lcccc}
\toprule
Selection Strategy & Avg \# DEGs & Pearson & Spearman \\
\midrule
Top-20 (like GEARS)           & 20                  & $0.614 \pm 0.066$ & $0.600 \pm 0.043$ \\
Top-100 (like scGen)          & 100                 & $0.594 \pm 0.024$ & $0.619 \pm 0.041$ \\
Strict (lfc$>$1, p$<$0.01)    & $15.3 \pm 5.1$      & $0.506 \pm 0.217$ & $0.370 \pm 0.182$ \\
Relaxed (lfc$>$0.25, p$<$0.1) & $71.7 \pm 6.9$      & $0.622 \pm 0.079$ & $0.609 \pm 0.048$ \\
\bottomrule
\end{tabular}
\end{table}

The choice between top-N and threshold-based selection involves a trade-off: top-N selection ensures consistent gene counts across conditions, enabling fair comparison, while threshold-based selection adapts to the biological signal strength of each perturbation. \geneval{} supports both strategies through the \texttt{n\_top\_degs} parameter in \texttt{DEGSettings}. Table \ref{tab:deg_thresholds} reports the results of our ablation study.

\section{Comparison with cell-eval}

The Arc Institute\footnote{\href{https://arcinstitute.org/}{https://arcinstitute.org/}} recently released \texttt{cell-eval} as part of the STATE framework ~\citep{adduri2025state}, representing an important step toward standardized evaluation of single-cell generative models. While \texttt{cell-eval} provides a comprehensive benchmarking pipeline tailored to the STATE ecosystem, it makes several design choices that trade flexibility and transparency for workflow specialization.

In contrast, \geneval{} is designed as a lightweight, model-agnostic evaluation layer that emphasizes (i) explicit control over computation space (raw, PCA, DEG-restricted), (ii) transparent metric configuration, and (iii) ease of integration into existing research pipelines. These differences reflect distinct design goals: \texttt{cell-eval} optimizes for a specific large-scale benchmarking workflow, whereas \geneval{} prioritizes generality and reproducibility across heterogeneous experimental setups.

A detailed, implementation-level comparison is provided in Appendix~\ref{app:cell_eval_comparison}. Table~\ref{app:tab:framework_comparison} summarizes the main design differences. We emphasize that this comparison is conceptual rather than an empirical head-to-head benchmark, which we leave for future work.

\subsection{Limitations and Future Directions}

\geneval{} focuses on static evaluation of generated samples against ground truth. Future work could extend to temporal evaluation (trajectory fidelity), multi-modal evaluation, and counterfactual evaluation where no ground truth exists. Additionally, the choice of biologically meaningful metrics remains an active research area; while perturbation-effect correlation captures one aspect of biological relevance, other metrics measuring pathway enrichment or regulatory network fidelity may provide complementary information.

Finally, beyond metrics alone, standardised benchmarking also requires standardised datasets and splits. Integrating \geneval{} with emerging community benchmarking suites (e.g., the curated perturbation datasets and predefined splits in \texttt{pertpy} \citep{heumos2025pertpy}) would enable fully reproducible, end-to-end evaluation pipelines, facilitating fair comparison across models and accelerating method development.

\subsection{Conclusion}

We have presented \geneval{}, a standardized framework for evaluating gene expression generative models. Through a comprehensive survey of the literature, we identified that current evaluation practices are heterogeneous to the point of incomparability. Through theoretical analysis, we showed why metric values depend on computation space. Through experiments, we demonstrated that these differences are substantial and that standardization is essential for fair comparison. \geneval{} provides a unified API with explicit configuration, hardware acceleration, and biologically-motivated DEG-space evaluation, enabling the field to move toward reproducible and comparable benchmarking.

\section*{Code Availability}
\geneval{} is available as an open-source Python package (\href{https://pypi.org/project/gge-eval/}{https://pypi.org/project/gge-eval/}), installable via \texttt{pip}, with the full source code hosted on GitHub (\href{https://github.com/AndreaRubbi/GGE}{https://github.com/AndreaRubbi/GGE}). Documentation is available at \href{https://docs.andrearubbi.com/gge}{https://docs.andrearubbi.com/gge}.


\bibliography{references}
\bibliographystyle{gen2_iclr2026_workshop}

\appendix

\section{Detailed landscape of generative models}
\label{app:gen_models_landscape}

\subsection{Generative Models for Single-Cell Data}
The landscape of generative models for single-cell gene expression has evolved
substantially over the past five years, with distinct architectural paradigms
emerging for different biological applications. Understanding this evolution
provides essential context for the evaluation challenges that \geneval{} addresses.

\paragraph{Variational Autoencoders (VAEs).}
VAEs established the foundation for single-cell generative modeling.
scVI~\citep{lopez2018deep} introduced deep generative models for single-cell data,
using a VAE architecture with negative binomial likelihood to model count data.
scGen~\citep{lotfollahi2019scgen} extended this approach to perturbation prediction
by learning latent arithmetic operations that map control cells to perturbed states.
CPA~\citep{lotfollahi2021compositional} further generalized to combinatorial perturbations
through disentangled representations of cell state and perturbation effects.
These VAE-based approaches typically evaluate using reconstruction MSE and $R^2$
between predicted and observed expression, computed either on all genes or a subset
of highly variable genes.

\paragraph{Graph Neural Networks (GNNs).}
GNNs brought biological knowledge into model architectures.
GEARS~\citep{roohani2024predicting} pioneered using gene regulatory networks as inductive biases,
enabling prediction of unseen multi-gene perturbation combinations. By passing messages along known
gene--gene interactions from STRING and co-expression networks, GEARS demonstrated that structure-aware
models generalize better to novel perturbations. However, GEARS introduced a new evaluation metric---the
fraction of DEGs correctly predicted in the top-20---that is not directly comparable to distributional
metrics used by other methods.

\paragraph{Optimal Transport (OT).}
OT methods provided principled frameworks for modeling cellular transformations.
CellOT~\citep{bunne2023learning} formulated perturbation response as an optimal transport problem,
learning a transport map that minimizes the 2-Wasserstein distance between control and perturbed
distributions. This represented a shift from predicting mean expression to modeling full distributions.
However, the reported ``Wasserstein distance'' values depend critically on whether computation occurs
in raw gene space, PCA space, or on a subset of genes.

\paragraph{Flow Matching.}
Flow matching has recently emerged as a powerful paradigm for modeling distribution-to-distribution
transformations. Conditional Flow Matching~\citep{lipman2022flow} and minibatch optimal transport
variants~\citep{tong2023conditional} learn continuous velocity fields that transport samples between
source and target distributions. Flow-based approaches explicitly model full conditional distributions
$p(x_1 \mid x_0, c)$ rather than only conditional means, making distributional metrics central to
evaluation. Recent work such as CellFlow~\citep{klein2025cellflow} emphasizes careful benchmarking and
protocol design, yet reported gains remain difficult to compare across studies due to non-standardized
evaluation pipelines.

\paragraph{Transformer-based foundation models.}
Transformer-based foundation models represent a complementary and increasingly influential direction.
STATE~\citep{adduri2025state} from the Arc Institute introduces a large-scale transformer architecture for
learning cellular dynamics and perturbation responses in a unified framework. Rather than learning explicit
transport maps or continuous flows, STATE models transitions between cellular states using attention-based
sequence modeling in learned latent spaces, enabling multi-task generalization across perturbations,
trajectories, and experimental contexts. These models emphasize scalability, representation learning, and
transfer across biological regimes, but introduce additional evaluation practices and metrics that are not
directly aligned with those used in VAE-, OT-, or flow-based approaches.

\section{Distributional Metrics: Formal Definitions and Computation}
\label{app:metrics_theory}

\subsection{Optimal Transport and Wasserstein Distances}

Optimal transport provides a principled framework for comparing distributions by computing
the minimum cost of transforming one distribution into another~\citep{villani2008optimal, peyre2019computational}.
The $p$-Wasserstein distance between distributions $P$ and $Q$ over $\mathbb{R}^d$ is defined as:
\begin{equation}
    W_p(P, Q) = \left( \inf_{\gamma \in \Gamma(P, Q)} \int \|x - y\|^p \, d\gamma(x, y) \right)^{1/p},
\end{equation}
where $\Gamma(P, Q)$ denotes the set of couplings with marginals $P$ and $Q$.

The 1-Wasserstein distance ($W_1$), also known as the Earth Mover's Distance, admits efficient
computation in low dimensions~\citep{rubner2000emd}. The 2-Wasserstein distance ($W_2$) is
particularly relevant for continuous-time generative modeling due to its connection to the
Benamou--Brenier formulation of optimal transport as a fluid dynamics problem~\citep{benamou2000computational_ot},
which underlies recent flow matching and diffusion-based models~\citep{lipman2022flow, tong2023conditional}.

For scalable computation, we employ entropy-regularized OT via the Sinkhorn algorithm~\citep{cuturi2013sinkhorn}:
\begin{equation}
    W_{p,\epsilon}(P, Q) =
    \left( \min_{\gamma \in \Gamma(P, Q)} 
    \int \|x - y\|^p \, d\gamma(x, y) + \epsilon H(\gamma) \right)^{1/p},
\end{equation}
where $H(\gamma)$ denotes the entropy of the coupling. The regularization parameter $\epsilon$
controls the bias--variance trade-off: larger $\epsilon$ yields smoother but more biased estimates,
while smaller $\epsilon$ approximates true OT at higher computational cost and sensitivity to noise.
In \geneval{}, we use $\epsilon=0.05$ by default, following common practice in large-scale OT-based
evaluation~\citep{feydy2019interpolating}.

\subsection{Maximum Mean Discrepancy}

Maximum Mean Discrepancy (MMD) measures distributional discrepancy via kernel mean embeddings
in a reproducing kernel Hilbert space~\citep{gretton2012kernel}:
\begin{equation}
    \mmd^2(P, Q) = 
    \mathbb{E}_{x,x' \sim P}[k(x,x')] 
    - 2\mathbb{E}_{x \sim P, y \sim Q}[k(x,y)] 
    + \mathbb{E}_{y,y' \sim Q}[k(y,y')],
\end{equation}
where $k$ is a characteristic positive-definite kernel. We use the Gaussian RBF kernel with
bandwidth selected via the median heuristic~\citep{gretton2012kernel}. MMD provides a consistent,
nonparametric two-sample test without explicit density estimation and is widely used for
benchmarking generative models~\citep{li2017mmdgan}.

\subsection{Energy Distance}

Energy distance provides an alternative metric grounded in statistical energy theory~\citep{szekely2013energy}:
\begin{equation}
    E(P, Q) = 
    2\mathbb{E}_{x \sim P, y \sim Q}[\|x - y\|] 
    - \mathbb{E}_{x,x' \sim P}[\|x - x'\|] 
    - \mathbb{E}_{y,y' \sim Q}[\|y - y'\|].
\end{equation}
Energy distance equals zero if and only if $P = Q$, making it a proper metric on probability
distributions. While it lacks the explicit geometric interpretation of OT, it provides a
computationally efficient and statistically well-founded measure of global distributional
similarity.

\section{Detailed Comparison with cell-eval}
\label{app:cell_eval_comparison}

We provide a more granular comparison between \geneval{} and the \texttt{cell-eval} framework~\citep{adduri2025state}, focusing on API design, configurability, extensibility, and documentation.

\paragraph{API Design and Workflow.}
\texttt{cell-eval} adopts a multi-stage CLI-based workflow involving explicit preprocessing, execution, baseline generation, and scoring steps. While this design supports reproducible large-scale benchmarking, it requires users to manage intermediate files, directory structures, and pipeline orchestration. By contrast, \geneval{} exposes a single high-level Python entry point that performs lazy data loading, per-condition metric computation, and aggregation within a unified interface, facilitating integration into custom training and evaluation loops.

\paragraph{Configuration Transparency.}
\texttt{cell-eval} relies on registry-based ``profiles'' (e.g.\ \texttt{full}, \texttt{minimal}, \texttt{vcc}) that bundle multiple metrics and preprocessing choices. While convenient for standardized benchmarks, these profiles obscure which metrics and spaces are being used without inspecting source code. In \geneval{}, all metric choices and hyperparameters are specified explicitly via typed metric objects with documented arguments.

\paragraph{Control over Computation Space.}
\texttt{cell-eval} computes metrics in a fixed representation space, without user-level control over PCA dimensionality or DEG selection thresholds. In contrast, \geneval{} treats representation space as a first-class parameter of each metric (e.g.\ raw gene space, PCA space with configurable dimensionality, or DEG-restricted space with tunable thresholds), enabling systematic sensitivity analyses.

\paragraph{Extensibility.}
Extending \texttt{cell-eval} with new metrics requires interacting with internal registries and enumerated types. \geneval{} exposes a minimal abstract metric interface, allowing custom metrics to be implemented with minimal boilerplate, lowering the barrier to methodological extensions.

\paragraph{Documentation and Usability.}
\texttt{cell-eval} primarily documents usage through CLI help text and examples embedded in the STATE codebase. \geneval{} provides a standalone documentation website with API references, tutorials, and reproducible examples.

\begin{table}[h]
\centering
\caption{Comparison of \geneval{} and \texttt{cell-eval} design choices.}
\label{app:tab:framework_comparison}
\vspace{0.5em}
\begin{tabular}{@{}lcc@{}}
\toprule
\textbf{Feature} & \textbf{GGE} & \textbf{cell-eval} \\
\midrule
Python API & \checkmark{} (single call) & \checkmark{} (multi-step) \\
CLI & \checkmark & \checkmark \\
Explicit space control & \checkmark & --- \\
Configurable DEG thresholds & \checkmark & --- \\
Top-N DEG selection & \checkmark & --- \\
GPU acceleration (CUDA) & \checkmark & --- \\
Apple MPS acceleration & \checkmark & --- \\
Extensive documentation & \checkmark & Limited \\
Simple custom metrics & \checkmark & Registry pattern \\
Visualization module & \checkmark & --- \\
\bottomrule
\end{tabular}
\end{table}

\section{Use of Large Language Models}
\label{app:llm_usage}

Large Language Models were used as writing assistants to improve clarity and presentation of the manuscript and software documentation. All scientific ideas, methods, experiments, and conclusions are solely those of the authors, and all LLM-assisted content was reviewed and edited for correctness prior to inclusion.

\end{document}